\documentstyle[pre,tighten,preprint,aps,amsmath,psfig]{revtex}

\renewcommand{\vec}[1]{\boldsymbol{#1}}
\newcommand{\mat}[1]{\boldsymbol{#1}}
\newcommand{\mcal}[1]{{\mathcal{#1}}}
\newcommand{\scalar}[2]{#1\odot #2}

\begin{document}

\title{Small world effects in evolution}
\author{Franco Bagnoli}
\address{Dipartimento di Matematica Applicata, Universit\`a di Firenze\\
Via S. Marta, 3 I-50139 Firenze, Italy\\
bagnoli@dma.unifi.it\\
also INFM sez. di
Firenze, Largo E. Fermi, 2 I-50125 Firenze, Italy
}
\author{Michele Bezzi}
\address{SISSA - Programme in Neuroscience\\
Via Beirut 2-4 I-34103, Trieste, Italy\\
bezzi@sissa.it\\
also INFM sez. di
Firenze, Largo E. Fermi, 2 I-50125 Firenze, Italy
}

\maketitle

\begin{abstract}
For asexual organisms point mutations correspond to local
displacements in the genotypic space, while other genotypic
 rearrangements
represent long-range jumps. 
We investigate the spreading 
properties of an initially homogeneous population in a flat fitness
landscape, and the equilibrium properties 
on a smooth fitness landscape. 
We show that a small-world effect is
present: even a small fraction of quenched long-range
jumps makes the results indistinguishable from those obtained by
assuming all mutations equiprobable. 
Moreover, we find that 
the equilibrium distribution is a Boltzmann one, in which the fitness
plays the role of an energy, and mutations that of a temperature.
\end{abstract}

\section{Introduction}
\label{sec:Introduction}

Darwinian evolution on asexual organisms acts by two mechanisms: 
{\it mutations}, that
increase the genetic diversity,  and {\it
selection}, that fixes and reduces this diversity. 

We  classify mutations into point ones, 
corresponding to local displacements on the genotypic space (defined
more accurately in the following), and other type of mutations or
rearrangements, which  imply larger jumps. Mutations are quite rare,
so we can safely assume that for each generation at most only one
mutation occurs.  
From a biological point of view, this discussion applies to 
unicellular asexual organisms, for which there is no 
distinction between somatic and germ cells. Moreover, 
we  do not consider the possibility of recombination 
(exchange of genetic material).  
For non-recombinant (asexual) organisms, the
combined effects of reproduction and mutations correspond to a random
walk on the genotypic space. Even sexual (recombinant) populations do
transmit asexually  some part of their genotype, such as
mithocondrial or y-chromosome DNA, to which the following analysis
applies. Furthermore, we assume a constant
environment.  

We schematize the effects of selection by the selective
reduction of the survival probability. 
Selection results from several constraints and  can
affect the frequency of certain genotypes, or even eliminate them 
from the population. We classify the components of selection  into
static and dynamic ones. In the first class we put the constraints
that are independent on the population
distribution, such as the functionality of a certain protein or the
tuning of a metabolic path. The static part of  the selection  is
equivalent to  the concept of ``fitness
landscape''~\cite{Fisher,Wright32}.  The dynamic part of the
selection originates from to the competition among individuals.  We
assume in this schematization that the competition only arises 
because  the population shares some external resource which is evenly
distributed, i.e. we disregard the effects of the inter-individual
competition~\cite{Bagnoli:prl,Bagnoli:review}.  In this limit the
effects of competition do not depend on the distribution of
genotypes, and simply limit the total size of the population.  

The most important effects of the selection can be roughly
schematized by assuming that certain genotypes are forbidden, so that
they are eliminated from the accessible space. Let us consider for the 
moment that evolution takes place in a flat fitness landscape.  In
this frame, there are no interactions among individuals:  the
evolution is given by  the simple superposition  of all possible
lineages.  The probability distribution of the population in the
genotypic space at a given time can be obtained by summing over all
possible individual ``histories'' in a way similar to the path
integral formulation of statistical mechanics~\cite{Bagnoli:path}. 
In this view, the phylogenetic lineage of an individual is given by
the path connecting the genotypes of ancestors. 

This assumption has to fail somewhere, since otherwise  evolution 
would be equivalent to  a diffusion process, without anything
favoring the formation of species.  It often assumed that speciation
events are rare and occur in a very short time~\cite{Gould}, either
due to some change in the fitness landscape (caused by an external
catastrophes like the fall of a large meteorite,  or by an internal
rearrangement induced by coevolution) or because small isolated
populations,  escaping from competition, are free to explore their
genotypic space (genetic drift) and find a path to an higher fitness
peak~\cite{Mayr}. Therefore, it is usually assumed that one can
reconstruct the diverging time of two species from their genotypic
distance from a single speciation event~\cite{phylogenetic}.

It is however known from recent investigations about the small-world
phenomenon~\cite{smallworld}, that diffusion is seriously affected
even by a small fraction of long-range jumps~\cite{Monasson}.  This
fact could have dramatic consequences in our understanding of the
large-scale evolution mechanism. If the short-range mutations are
dominant, the evolution is equivalent to  a diffusion process in the
genotypic space. Assuming that the fitness landscape is formed by
mountains separated by valleys, and that the crests of mountains are
almost flat, then the large-scale evolution is dominated by the times
needed to cross a valley by chance, while the short-scale is
dominated by the neutral exploration of crests~\cite{Gould,Kimura}.

Vice versa, if the long-range mutations are important (eventually
amplified by the small-world mechanism), the time needed to connect
two genotypes does not depend on their distance nor by the shape of
the fitness landscape and the fitness maxima are quickly populated.
Moreover, in this scenario, the speciation phenomenon should not be
ascribed to the ``discovery'' of a preexisting niche, but rather to
the formation  of that niche in a given
ecosystem due to internal interactions (coevolution) or external 
physical changes (catastrophes). After the formation, the niche is quickly 
populated because of the long-range mutations. 
In this framework, allopatric
speciation~\cite{Mayr} looses its fundamental importance, and
sympatric speciation due to coevolution~\cite{Kondrashov,Dieckmann}
becomes a plausible alternative. 

In order to evaluate quantitatively the relevance of these
speculations, we introduce an individual-based model of an evolving
population. We assume that the
genetic information (the genotype) of an individual is represented by
a binary string $\vec{x}=(x_1, x_2, \dots, x_L)$ of $L$ symbols $x_i=0,1$
(multilocus model  with two alleles). 
 In this way we are modeling
haploid organisms, i.e.\  bacteria or viruses or, more appropriately,
more archaic, pre-biotic entities. The choice of a binary code is not
fundamental but certainly makes things easier. It can be justified by
thinking at a purine-pyrimidine coding, or to ``good'' and ``bad''
alleles for genes (units of genetic information). In this second
version, $0$ represents a good gene and $1$ a bad one. 
This is a sort of ``minimal model'' which is often used to 
model the evolution of genetic 
populations~\cite{Kondrashov,Dieckmann,Eigen,Doebeli,speciation}.
For bacteria,
one can give indications that major and minor codons work in such a
way~\cite{BagnoliLio}.  The genotypic space is thus a Boolean
hypercube of $L$ dimensions, and each individuals sits on a corner of
this cube, according to its  genotype. The $2^L$ corners 
of this hypercube represent all
possible genotypes, which are
at a maximum distance equal to $L$. 

In the next section,  we formalize the model and  introduce  the
mathematical representation of mutation mechanisms. Then, in
Section~\ref{sec:flat}, we consider the consequences of short and
long-range mutations for a  flat fitness  landscape.   In the
long-range case all genotypes are connected, regardless of their mutual
distance: this case can be considered the equivalent of a mean-field
approximation. We derive analytically the rate of spreading $v$ and
the characteristic spreading time $\tau$ of a genetically homogeneous
inoculum in the short-range ($v_s$, $\tau_s$) and long-range
($v_{\ell}$, $\tau_{\ell}$) cases. 
In the first case, $v_s$ is
independent of the genotype length $L$, and $\tau_s$ grows linearly
with $L$; the opposite happens in the long-range case. Clearly, this
different behavior has dramatic evolutionary consequences  as $L$
becomes large.

In general, however, only a small set of all possible long-range
mutations is observed in real organisms. We thus compute numerically
the rate of spreading $v$ in a mixed short-range and sparse
long-range case. A small-world effect can be observed: in the limit of
large genotype lengths, a vanishing fraction of long-range mutations
cooperates with the short-range ones to give essentially the
mean-field results.

In Section~\ref{sec:path}, we show that  the effects of  mutations
and  selection can be separated in the limit of a very smooth fitness
landscape and mean-field long-range mutations. In this approximation
we obtain analytically that  the asymptotic probability distribution
is a Boltzmann equilibrium one, in which the fitness plays the role
of of an energy and mutations correspond to temperature.  We compute
numerically the asymptotic probability distribution for several
fitness landscapes, finding that the  equilibrium hypothesis is
verified. As a nontrivial example of small-world effects in
evolution, we checked numerically that this scenario still holds for
a sparse long-range mutation matrix. Conclusions and perspectives are
drawn in the last section.

\section{The model}
\label{sec:model}

In order to be specific, let us assume that the genotype of an individual
is represented by  a Boolean
string of length $L$. In this way, the genotypic space is a Boolean
hypercube with $2^L$ nodes. 

The genotype of an
individual is represented by a  string $\vec{x}=( x_1, x_2, \dots, x_L)$
of $L$ Boolean symbols $x_i=0,1$. 
Each position corresponds to a locus whose gene has two allelic forms. 
In this way we are modeling haploid (only
one copy of each gene) organisms, i.e.\  bacteria or viruses or, more
appropriately, more archaic, pre-biotic entities.
The genotype can be also viewed as a spin configuration. In this case we
use the symbol $\sigma_i = 2x_i -1$. 

We  consider two kinds of mutations: point mutations, that interchange
a $0$ with a $1$, and all other mutations that do not alter
 the length of
the genotype (transposition and inversions).
We define the distance $d(\vec{x},\vec{y})$ between two genotypes $\vec{x}$ and $\vec{y}$
as the minimal number of point mutations needed to pass from $\vec{x}$ to
$\vec{y}$ (Hamming distance) 
\[
	d(\vec{x},\vec{y}) = \sum_{i=1}^L (x_i-y_i)^2.
\]
All possible genotypes of length
$L$ are distributed on the $2^L$ vertices of an hypercube. 
A point mutation
corresponds to a unit displacement on that hypercube (short-range
jump).

The occurrence of point mutations in real organisms 
 depends on the identity of the symbol and 
on its position  on the genotype; in the present approximation, however, 
we assume that
all point mutations are equally likely. Moreover,  since the probability of
observing a mutation is quite small, we impose that at most one
mutation is possible in one generation.
The probability to observe a point mutation from genotype $\vec{y}$ to genotype $\vec{x}$  
is given by
the short-range mutation matrix $M_s(\vec{x},\vec{y})$.
By denoting with $\mu_s$ the probability
of a point mutation per generation, we have 
\begin{equation}\label{Ms}
	M_s(\vec{x},\vec{y}) = 
  \left\{\begin{array}{ll}
	  1-\mu_s & \mbox{if $\vec{x}=\vec{y}$,} \\
		  \dfrac{\mu_s}{L}& \mbox{if $d(\vec{x},\vec{y})=1$,} \\
	  0&\mbox{otherwise.}
  \end{array}\right.
\end{equation}
 
Other types of mutations correspond to long-range jumps in the genotypic space.
The simplest approximation
 consists in assuming all mutations
equiprobable. Let us denote with $\mu_{\ell}$ the probability per
generation of this kind of mutations. The long-range mutation matrix,
$M_{\ell}(\vec{x},\vec{y})$, is defined as
\begin{equation}\label{Ml}
	M_{\ell}(\vec{x},\vec{y}) = 
	\left\{\begin{array}{ll}
		 1-\mu_{\ell} & \mbox{if $\vec{x}=\vec{y}$,} \\
		\dfrac{\mu_{\ell}}{2^L-1}&\mbox{otherwise.}
	\end{array}\right.
\end{equation}
In the real world, only some kinds of mutations are actually observed. We model this fact 
by replacing $\vec{M}_{\ell}$ with a sparse matrix $\hat{\mat{M}}_{\ell}$. 
We introduce the sparseness index
$s$, which is the average number of nonzero off-diagonal elements of
$\hat{\mat{M}}_{\ell}$. The sum of these off-diagonal elements 
still gives $\mu_{\ell}$. In this case $\hat{\mat{M}}_{\ell}$ 
is a quenched sparse
matrix, and $\mat{M}_{\ell}$ can be 
considered the average of the annealed version.

After considering both types of mutations, the overall mutation matrix
is $\mat{M} =
\mat{M}_{\ell}\mat{M}_s$ or $\mat{M} =
\hat{\mat{M}}_{\ell}\mat{M}_s$ for the quenched version.

We model our population at the level of the probability distribution of genotypes
$\vec{p}\equiv\vec{p}(t)$, thus disregarding spatial
effects.  The evolution equation for
$\vec{p}$ is 
\begin{equation}\label{mutsel}
	p'(\vec{x})=
	\dfrac{
	A(\vec{x})\sum_{\vec{y}} M(\vec{x},\vec{y})p(\vec{y})}{\overline A},
\end{equation}
where the selection function $A(\vec{x})$ corresponds
to the average reproduction
rate of individuals with genotype $\vec{x}$, and  
${\overline A}= \sum_{\vec{x}} A(\vec{x}) p(\vec{x})$
is the average reproduction rate of the 
population~\cite{Fisher,Wright32,Hartle,Peliti95}.
We write $A(\vec{x})$ in an exponential form $A(\vec{x})=\exp(V(\vec{x}))$, and we denote
$V(\vec{x})$ with the term \emph{fitness landscape}.

The selection does not act directly on the genotype, but rather on 
the {\it phenotype} (how an individual
appears to others). The phenotype of a given genotype 
can be interpreted as an array of
morphological characteristics. We consider the simplest case, in
which  the phenotype if univocally 
determined by  the genotype $\vec{x}$, which is not the
general case, since polymorphism or age dependence are usually
present. The general mapping between genotype and phenotype is largely
unknown and is expected to be quite complex. The effects of some genes
are additive (non-epistatic), while others can interact in a simple
(control genes) or complex (morphologic genes) way.

A possible way of approximating these effects is to 
use the following form for the fitness $V(\vec{x})$:
\begin{equation}
\label{fitness}
	V(\vec{x}) = \dfrac{\mcal{H}}{L} \sum_{i=1}^L \sigma_i + 
		\dfrac{\mcal{J}}{L-1} \sum_{i=1}^{L-1}
		\sigma_i \sigma_{i+1} + \mcal{K} \eta(\vec{x}),
\end{equation}
where $\eta(\vec{x})$ is a random function of $\vec{x}$, uniformly distributed between
$-1$ and $1$ ($\langle \eta(\vec{x}) \eta(\vec{y}) \rangle = \delta_{\vec{x}\vec{y}}$). 
The ``field'' term $\mcal{H}$ represents  a non-epistatic
contribution to the fitness, in which all genes have equal weight. 
The ``ferromagnetic'' term $\mcal{J}$ represents simple
interactions between pairs of genes (even though in general those are
not symmetric) and corresponds to 
 a weakly rough landscape. Finally, 
 $\mcal{K}$ modulates a widely rough landscape and can
be thought as an approximation of the effects of complex  (spin-glass
like) interactions among genes.

\section{Spreading and small-world effects on a flat fitness landscape}
\label{sec:flat}
In this section we study the case of a flat fitness landscape (no
selection), i.e.\ $\mcal{H}=\mcal{J}=\mcal{K}=0$. 
In this way we are modeling the evolution on the crest of a 
mountain, assuming that all deleterious mutations are immediately
lethal (the part of genotype that can originate this kind of mutations
is not considered), and that we can neglect the small variation of
fitness along the
crest. This landscape is the one usually considered in  the
theory of neutral evolution~\cite{Kimura}. Let us assume that this
crest is colonized by a founder deme genotypically homogeneous (a
delta peak in the genotypic distribution). We want to obtain
the average time needed to populate the crest according with the different
mutation schemes. Since  the fitness landscape is flat, $\overline{A}$
is a constant. 

We are interested in the behavior of $\rho(t)$, 
which is the average
distance of the population from a given starting genotype $\vec{x}_0$, i.e.
\begin{equation}
	\rho(t) = \sum_{\vec{x}} p(\vec{x},t) d(\vec{x},\vec{x}_0). \label{rho}
\end{equation}
We introduce the spreading velocity $v$ as 
\[
	v = \left.\dfrac{\partial \rho}{\partial t}\right|_{t=0}. 
\]

In Appendix A we obtain the spectral properties of the mutational
matrices $\mat{M}_s$ and $\mat{M}_{\ell}$. The corresponding
spreading velocities  in the limit of small time interval $t$
compared to the characteristic spreading time $\tau_s = L/2\mu_s$
(short-range) and $\tau_{\ell}=1/\mu_{\ell}$ (long-range) are $v_s =
\mu_s$, Eq.~(\ref{vs}), and $v_{\ell} = L\mu_{\ell}/2$,
Eq.~(\ref{vl}).  If short and long-range mutations coexists, one has 
$v = v_s + v_\ell$, Eq.~(\ref{v}).

One can approximate the behavior of real biological systems by
considering a mixture of short-range mutations, which occur with a
relatively high frequency, and sparse long-range mutations, with
sparseness index $s$. 

\begin{figure}
\psfig{figure=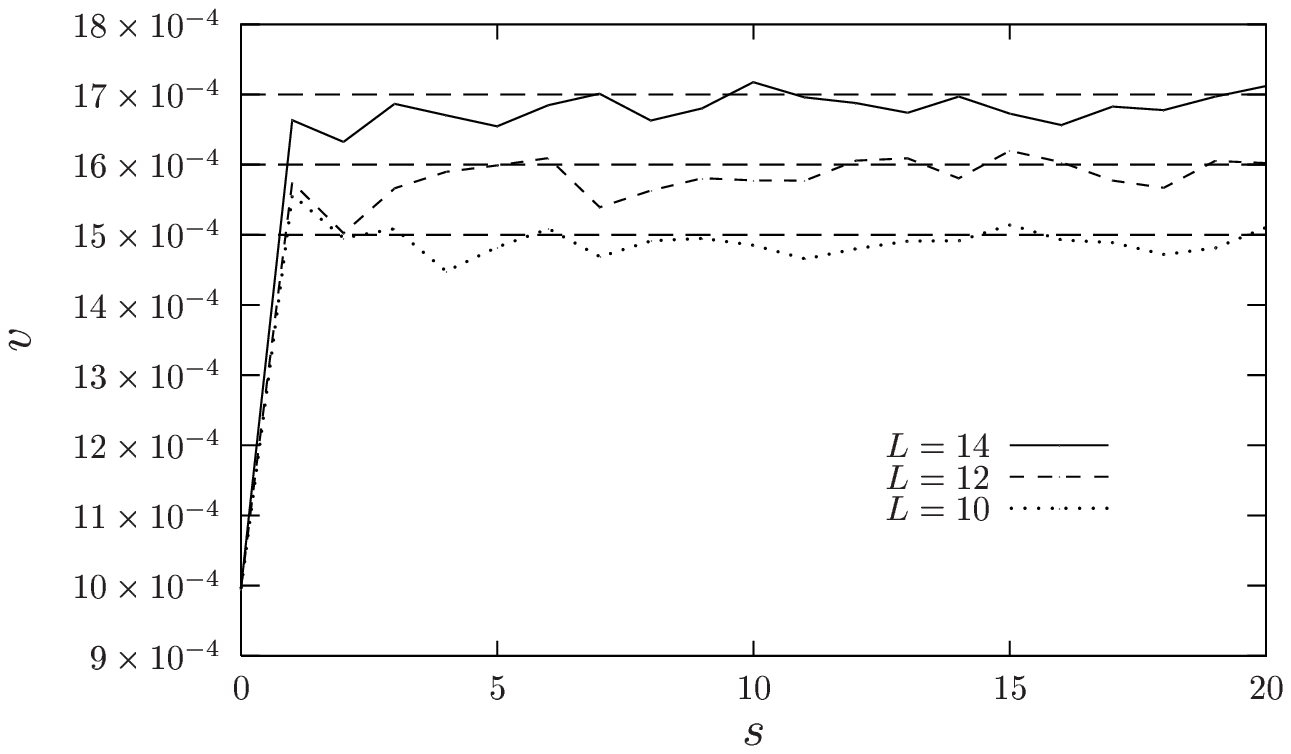,width=12cm} 
\caption{
The spreading velocity $v$ vs.\ sparseness $s$ for three
values of the genotype length $L$; $\mu_{\ell}=10^{-4}$, $\mu_s=10^{-3}$,
flat fitness landscape. The dashed lines indicate the 
mean field value $v=\mu_s+L\mu_{\ell}/2$.
	\label{figure:spreading}
	}
\end{figure}

We investigated the sparse case by numerical simulations, for some
genotype lengths $L$.  As shown in Figure~\ref{figure:spreading},  as
soon as the sparseness index $s>0$, the numerical value of $v$
becomes very close to the mean-field one, Eq.~\eqref{v}.  Notice
that the average distance from inoculum, $\rho(t)$, is  rather
insensitive to the distribution of $\vec{p}$. The actual
distribution  can be quite different from the one obtained with the
mean-field matrix $\mat{M}_{\ell}$. 

\begin{figure}
\psfig{figure=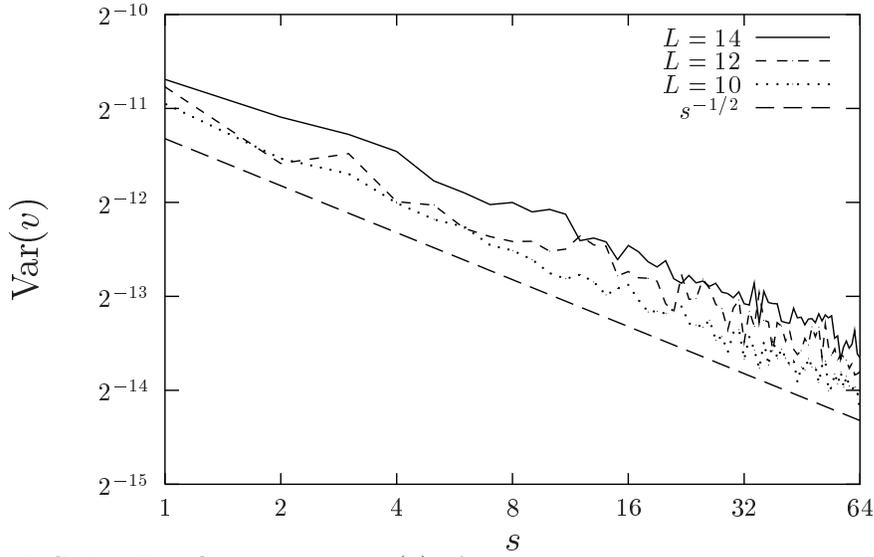,width=12cm} 
\caption{
The fluctuations $\text{Var}(v)$ of the spreading velocity $v$
vs.\ sparseness $s$  for for three values of the 
genotype length $L$;
$\mu_{\ell}=10^{-4}$, $\mu_s=10^{-3}$,
flat fitness landscape. The dashed line represent the law
$\text{Var}(v)\sim s^{-1/2}$ .
	\label{figure:fluctuations}
}
\end{figure}

This transition may be interpreted as an indication of a small-world 
effect, and that there exists a 
first-order transition at $s=0$~\cite{first-order}.
However, the fluctuations of  the velocity, $\text{Var}(v)$, appears to
diverge at $s=0$ with an exponent $1$, as shown in
Figure~\ref{figure:fluctuations}. 

These results suggest the presence of a  small-world phenomenon in
evolution: the rare and sparse long-range mutations cooperate
synergetically with the short-range ones to give essentially a
mean-field effect. We checked that the previous results hold also for
a non-flat (but smooth) fitness landscape ($\mu_s =10^{-5}$,
$\mu_{\ell}=10^{-4}$ and $\mcal{H}=10^{-4}$, $\mcal{J}=0$,
$\mcal{K}=0$). Again, as soon as $s>0$, the spreading velocity
increases from the short-range values to the mean-field ones.

\section{Equilibrium properties and small-world effects on a smooth fitness landscapes}
\label{sec:path}

In the following we shall study the effects of the cooperation between
short-range and sparse long-range mutations on the 
equilibrium properties of the model.

Let us study the model 
in the  presence of a smooth static fitness landscape. In this case
the fitness $A$ does not depend
explicitly on the distribution $\boldsymbol{p}$, and
Eq.~(\ref{mutsel}) can be linearized by using unnormalized variables 
$z(\vec{x},t)$ satisfying
\begin{equation}
	z(\vec{x}, t+1) = \sum_{\vec{y}} A(\vec{y}) M(\vec{x},\vec{y})z(\vec{y},t), \label{z}
\end{equation}
with the correspondence
\[
	p(\vec{x},t) = \dfrac{z(\vec{x},t)}{\sum_{\vec{y}}z(\vec{y},t)}.
\]
In vectorial terms, Eq.~(\ref{z}) can be written as 
\begin{equation}
	\vec{z}(t+1) = \mat{M}\mat{A}\vec{z}(t),
	\label{Z}
\end{equation}
where $\mat{M}_{\vec{xy}}=M(\vec{x},\vec{y})$ 
and $\mat{A}_{\vec{xy}}=A(\vec{x})\delta_{\vec{xy}}$.

When one takes into consideration only point mutations
($\mat{M}\equiv \mat{M}_s$), Eq.~(\ref{z}) can be read
as  the transfer matrix of a two-dimensional Ising
model~\cite{Leuthausser,Tarazona,Baake},  for which the genotypic
element $\sigma_i(t)$ corresponds to the spin in row $t$ and column
$i$, and $ z(\vec{\sigma},t)$ is the restricted partition function of row
$t$.  The effective Hamiltonian $\mcal{V}$ (up to  constant terms)  of a
possible genealogical history $\{\vec{x}(t)\}$ or $\{\vec{\sigma}(t)\}$ 
from time $1\le t \le T$ is
\begin{equation}
	\mcal{V} = \sum_{t=1}^{T-1} \left(\gamma \sum_{i=1}^L
	\sigma_i(t)\sigma_{i}(t+1) + V\bigl(\vec{x}(t)\bigr)\right),
	\label{ising} 
\end{equation}
where $\gamma=-\ln(\mu_s/(1-\mu_s))$.
 
This peculiar two-dimensional Ising model has a long-range coupling
along the row (depending on the choice of the fitness function) and a
ferromagnetic coupling along the time direction (for small short
range mutation  probability).  In order to obtain the statistical
properties of the system one has to sum over all possible
configurations (stories), eventually selecting the right boundary
conditions at time $t=1$.  The bulk properties of Eq.~(\ref{ising}) 
cannot be reduced in general to the equilibrium distribution of a one
dimensional system, since the transition probabilities among rows do
not obey detailed balance. Moreover, the temperature-dependent
Hamiltonian (\ref{ising}) does not allow an easy identification
between energy and selection, and temperature and mutation, what is
naively expected by the biological analogy with an adaptive walk. 

\subsection{Long-range mutations}

Let us first consider the long-range mutation case. 
Eq.~(\ref{Z}), reformulated according to Eq.~(\ref{mutsel}), corresponds to 
\[
	\vec{z}(t+\tau) = (\mat{A}\mat{M}_{\ell})^\tau\vec{z}(t).
\]
Since it is easier to consider the effects of 
$\mat{A}$ and $\mat{M}_{\ell}$ separately, let us 
study in which limit they commute.  
The norm of the commutator on the asymptotic probability distribution
$\vec{p}$ is 
\[
||[\mat{A}\mat{M}_{\ell}]||=\sum_{\vec{xy}}
|[\mat{A}\mat{M}_{\ell}]_{\vec{xy}}p(\vec{y})|,
\]
and it is bounded by $\mu_{\ell}c$, where $c= \max_{\vec{xy}}
|A_{\vec{xx}}-A_{\vec{yy}}|$.
In the limit $\mu_{\ell}c\rightarrow 0$ (i.e.\ a very smooth landscape), 
to first order in $c$ we have
\[
	(\mat{A}\mat{M}_{\ell})^\tau = \mat{A}^\tau\mat{M}_{\ell}^\tau +
	O(\tau^2 \mu_{\ell} c) \mat{A}^{\tau-1}\mat{M}_{\ell}^{\tau-1}, 
\]
which is the analogous of the Trotter product formula. 

When $\tau$ is order $1/\mu_{\ell}$, $\mat{M}_{\ell}^\tau$ is a constant
matrix with elements equal to $1/2^L$, and thus $\mat{M}_{\ell}\vec{p}$ is a 
constant probability distribution. If $\mu_{\ell}$ is large enough, 
$(\mat{A}\mat{M}_{\ell})^\tau = \mat{A}^\tau\mat{M}_{\ell}^\tau$.

The asymptotic probability distribution  
$\tilde{\vec{p}}$
is thus proportional to the diagonal of $\mat{A}^{1/\mu_{\ell}}$:
\begin{equation}
	\tilde p(\vec{x}) = C \exp\left(\dfrac{V(\vec{x})}{\mu_{\ell}}\right)
	\label{Boltzmann}
\end{equation}
i.e.\ a Boltzmann distribution with Hamiltonian $V(\vec{x})$ 
and temperature $\mu_{\ell}$. This corresponds to the naive analogy between
evolution and equilibrium statistical mechanics. In other words, the
genotypic distribution is equally populated if the phenotype is the
same, regardless of the genotypic distance since we used long-range
mutations.

We have checked numerically this hypothesis, by 
iterating Eq.~(\ref{mutsel}) for a time
$T$ large enough to be sure of having reached the asymptotic state. 
We plotted the logarithm of the probability
distribution $\tilde p(\vec{x})$ versus the value of the Hamiltonian
$V(\vec{x})$. We computed the slope $1/\mu_{e}$ of the linear regression.
The quantity $\mu_e$ is the effective ``temperature'' of the
probability distribution according to the equilibrium hypothesis.
 
\begin{figure}
	\psfig{figure=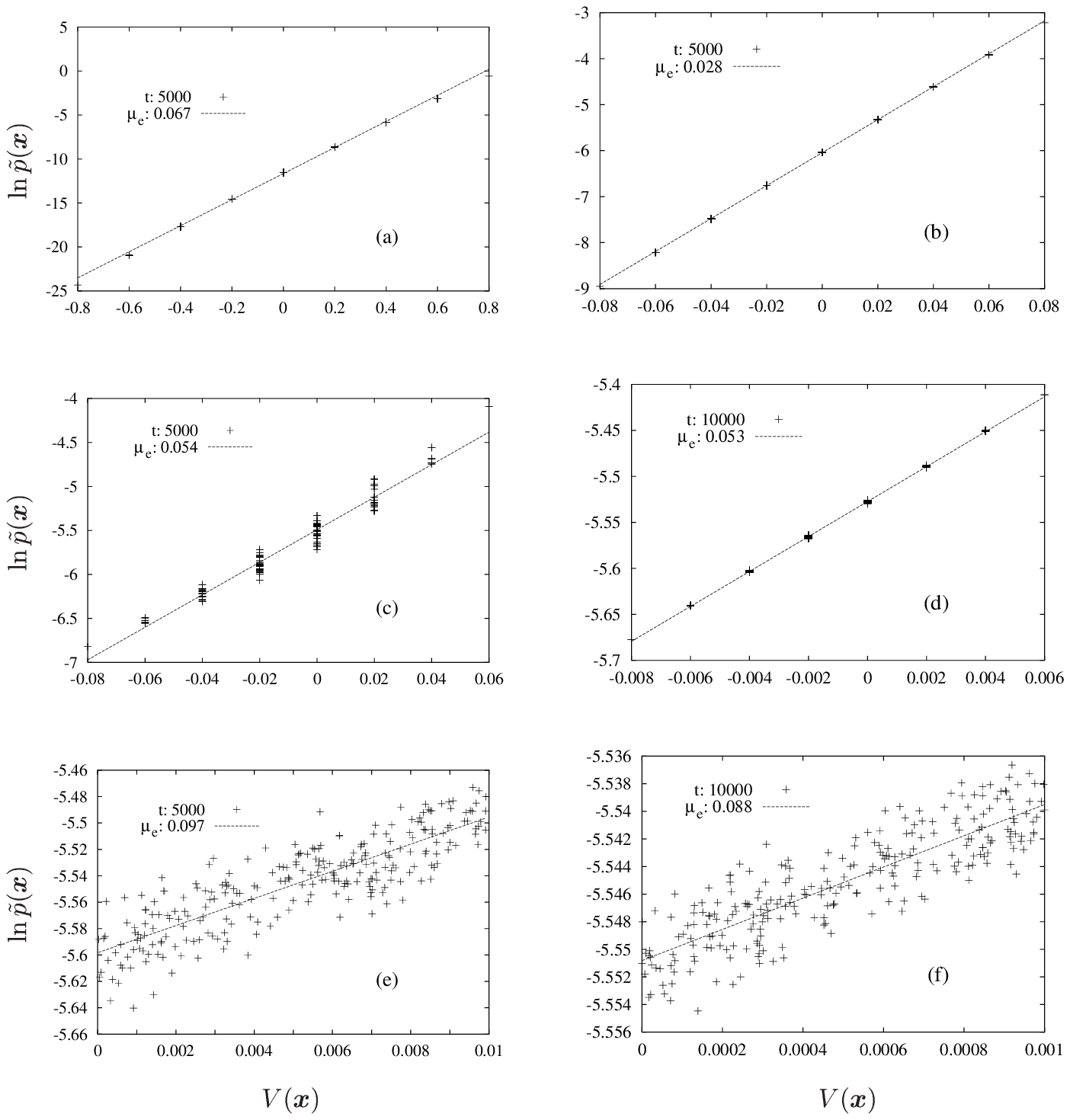}
	\caption{
	Numerical check for long-range mutations. In the simulations we set
	$L=8$, $\mu_{\ell}=0.1$ and  $\mu_s=0$.
	We varied  $\mcal{H}$ (a,b), $\mcal{J}$ (c,d) and $\mcal{K}$ (e,f),
	setting all other parameters to zero.
	(a)  $\mcal{H}=0.01$, $\mcal{J}=0$, $\mcal{K}=0$;
	(b)  $\mcal{H}=0.001$, $\mcal{J}=0$, $\mcal{K}=0$;
	(c) $\mcal{H}=0$, $\mcal{J}=0.01$, $\mcal{K}=0$;
	(d) $\mcal{H}=0$, $\mcal{J}=0.001$, $\mcal{K}=0$;
	(e) $\mcal{H}=0$, $\mcal{J}=0$, $\mcal{K}=0.1$;  
	(f) $\mcal{H}=0$, $\mcal{J}=0$, $\mcal{K}=0.01$. In the figures $t$
	indicates the number of generations, $\mu_{e}$ the reciprocal of the slope of
	linear regression.   
	\label{figure:Ml}
	}
\end{figure}

The results for the mean-field mutation matrix $\mat{M}_{\ell}$ 
are shown in Figure~\ref{figure:Ml}.
We see that the equilibrium hypothesis is well verified in the limit
$\mu_{\ell} \gg c$; and that convergence is faster for a rough landscape. 
In all cases the effective temperature $\mu_e$ 
is close to the expected value, i.e.\ to the mutation rate $\mu_{\ell}$.

\subsection{Short-range mutations}

\begin{figure}
	\psfig{figure=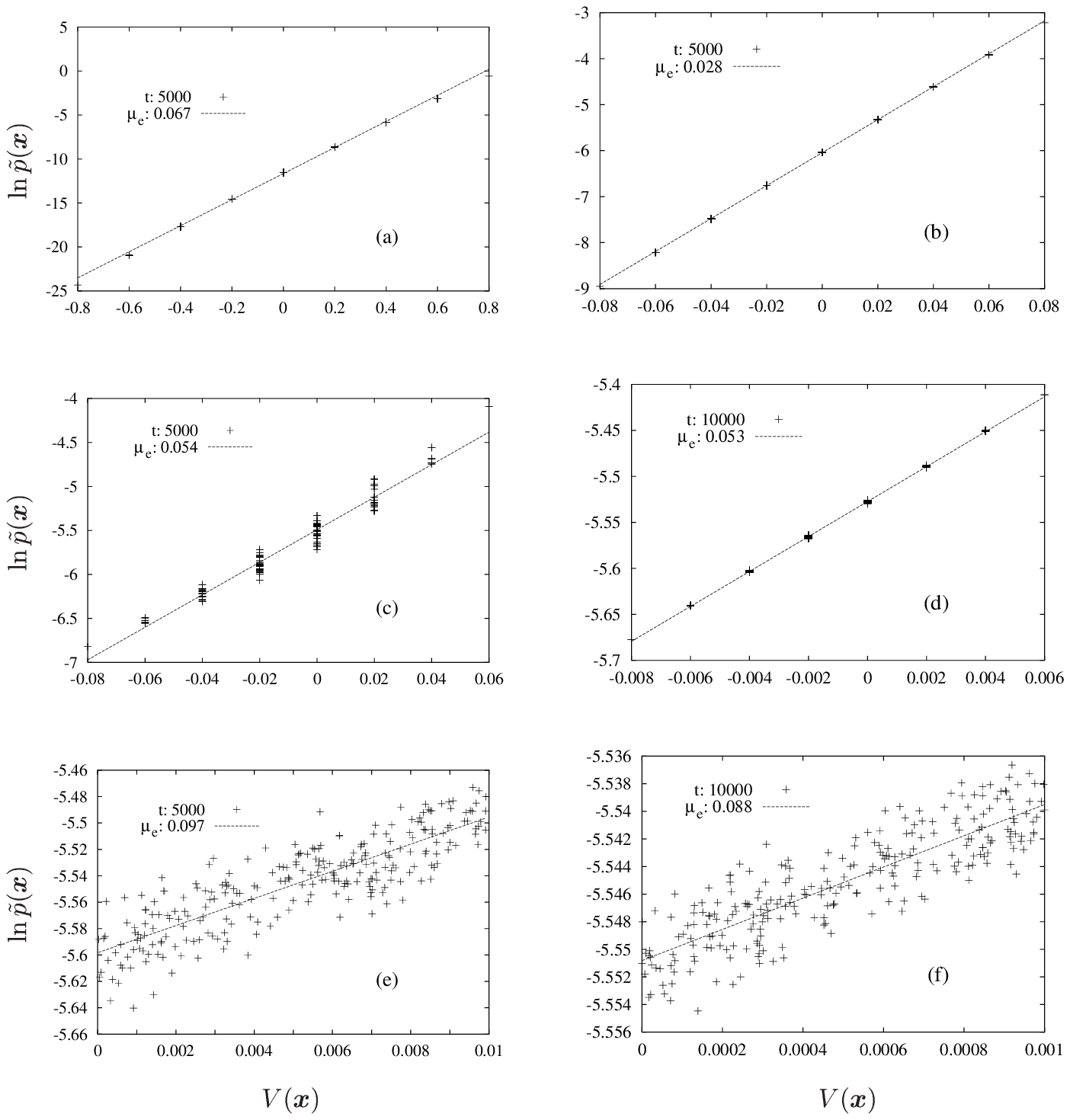}
	\caption{
	Numerical check for short-range mutations. In the simulations we set
	$L=8$, $\mu_{\ell}=0$ and  $\mu_s=0.1$.
	We varied  $\mcal{H}$ (a,b), $\mcal{J}$ (c,d) and $\mcal{K}$ (e,f),
	setting all other parameters to zero.
	(a)  $\mcal{H}=0.1$, $\mcal{J}=0$, $\mcal{K}=0$;
	(b)  $\mcal{H}=0.01$, $\mcal{J}=0$, $\mcal{K}=0$;
	(c) $\mcal{H}=0$,  $\mcal{J}=0.01$,$\mcal{K}=0$;
	(d) $\mcal{H}=0$, $\mcal{J}=0.001$, $\mcal{K}=0$;
	(e) $\mcal{H}=0$, $\mcal{J}=0$, $\mcal{K}=0.01$;  
	(f) $\mcal{H}=0$, $\mcal{J}=0$, $\mcal{K}=0.001$. 
	In the figures $t$
	indicates the number of generations, $\mu_{e}$ the reciprocal of the slope of
	linear regression.   
	\label{figure:Ms}
	}
\end{figure}

The above results only hold qualitatively for pure short-range mutations
as shown in Figure~\ref{figure:Ms}. A small dispersion of points in
figure implies that the genotypes can be divided into evenly-populated groups
sharing approximately the same fitness.
This is always the case for the additive fitness landscape 
($\mcal{H}$ contribution), since in this case the position of 
symbols in the genotype is ininfluent, and the short-range mutations are
able to homogenize the distribution inside a group. This is only approximately
valid for the $\mcal{J}$ contributions, since in this case the fitness depends 
on pairs of symbols, while mutations only act on single symbols. However, 
as shown in Figure~\ref{figure:Ms}-d, the homogeneous state is reached 
for a sufficiently high mutation probability. 
Finally, the homogeneous state  is never reached
 for the very rough landscape case ($\mcal{H}$ contribution), 
even though the linear relation is satisfied
in average. 

In order to obtain a  quantitatively correct prediction, one has to
consider that the resulting slope $\mu_{e}$ is related to the second
largest eigenvalue $\lambda_1$ of the mutation matrix by $\mu=1-\lambda_1$. 
When the fitness only
depends on the ``external field'' term $\mcal{H}$, in the asymptotic
state the short-range mutations connects group of equal fitness. 
This fact is reflected by the vanishing dispersion of points in Figure~\ref{figure:Ms}-a
and \ref{figure:Ms}-b. 

In all cases, we expect that in the limit of a very smooth landscape, 
$\mu_{e}$ tends towards the limiting value $2\mu_s/L$ of 
Eq.~(\ref{Msspectrum}). This limit is reached faster 
when only the $\mcal{H}$ term  is present. 
This implies that for large genotype, 
the effective temperature due to short-range mutations is vanishing. 

In the opposite case, when the selection is strong,  the application
of the matrix $\boldsymbol{A}$ ``rotates'' the distribution
$\boldsymbol{p}$ in a way which is practically random with respect to
the Fourier eigenvectors of $\boldsymbol{M}_s$. Thus, the effective
second eigenvalue of the mutation matrix is given by $1-\mu_s$, 
obtained averaging over all the eigenvalues of
Eq~(\ref{Msspectrum}).  Consequently, we obtain  $\mu_{e} \simeq
\mu_s$. This limit is reached faster in the case of a strongly disordered
fitness landscape,  i.e.\ in the limit of large $\mcal{K}$.  When
only the $\mcal{J}$ term is present, one observes an intermediate
case. 

\subsection{Small-world effects}

Let us now consider the case of a sparse long-range mutation matrix, coupled to 
a stronger short-range matrix, i.e.\ $\mu_s \gg \mu_{\ell}$. 
We performed simulations with $\mu_s=0.1$ and $\mu_{\ell}=0.01$, 
$L=6$, 8 and 10 and varying $s$. The results are summarized in Table~\ref{table:s}.

The effects of the two kinds of mutations is additive, implying that, for small
  $\mu_{\ell}$,  
the distributions are visually similar to the short-range case, 
Figure~\ref{figure:Ms}. However, as soon as $s$ is greater than a threshold 
$s_c$, the effective temperature $\mu_e$  increases by the expected 
long-range contribution, $\mu_{\ell}$. This increment is very relevant 
when the effect of short-range mutations are vanishing, i.e.\ for smooth
landscapes ($\mcal{H}$ and $\mcal{J}$ contributions) and large genotypes. 
The effect is less relevant in the disordered case,  
($\mcal{K}$ contributions), since in this case the contribution to the effective
 temperature by the small-range mutations do not decrease with $L$. 

By increasing the weigth of the sparse long-range mutations, also the probability 
distribution becomes similar to the long-range case, Figure~\ref{figure:Ml}. 
We performed simulations with $\mu_s=0.1$ and $\mu_{\ell}=0.1$, 
$L=6,8,10$ varying $s$. The results are shown in Table~\ref{table:s}. 
Notice that the value of $s_c$ (estimated visually from the plot of data) is rather 
insensitive to $L$ and the parameters of the fitness $V(\vec{x})$. 

\begin{figure}
	\psfig{figure=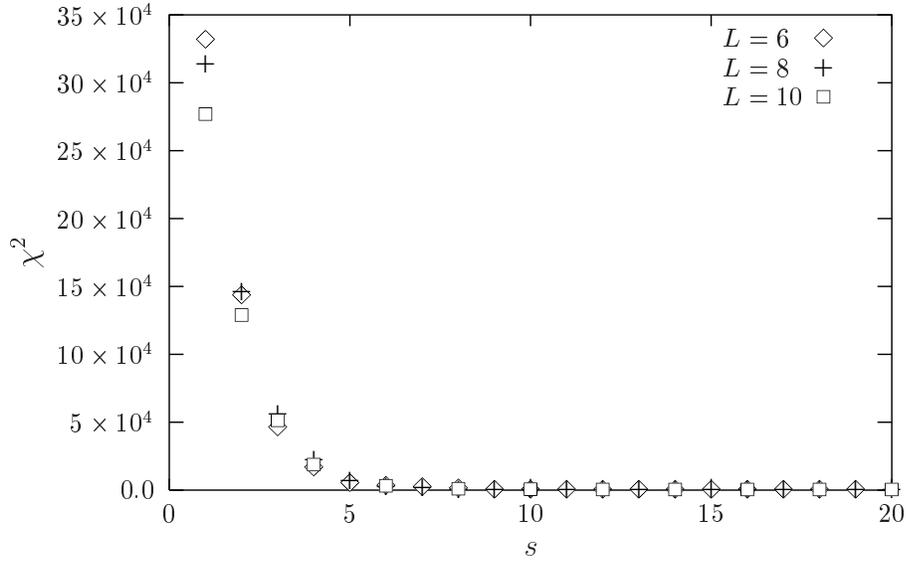,width=12cm}
	\caption{Scaling of the average square difference from linear regression, 
	$\chi^2$ vs.\ sparseness $s$, for
	three values of the genotype length $L$; $\mu_{\ell}=0.1$, $\mu_s=0.1$,
    $\mcal{H}=0$, $\mcal{J}=0$, $\mcal{K}=0.1$. Averages taken over 20 runs.} 
	\label{figure:sparsechi} 
\end{figure}

We quantified the distance between the resulted distribution 
 and the Boltzmann one
Eq.~(\ref{Boltzmann}) through the  computation of  the 
average square difference from linear regression, 
 $\chi^2$.
 In Figure~\ref{figure:sparsechi} we show that $\chi^2$ becomes very small 
 as soon as $s>s_c$, which seems to vanish with $L\rightarrow \infty$.
 
This implies that, for large genotypes,  
a vanishing fraction of long-range mutations, coupled to small-range ones,  
are sufficient to establish  the statistical mechanics
analogy of selection and mutations.

\section{Conclusions}
\label{CA:disc}
We studied some simple models of asexual populations evolving on a smooth
fitness landscape, in the presence of point mutations (small-range jumps in
genotypic space) and other genetic rearrangements (long-range jumps). 

We have computed analitically 
the spreading velocity of an initially homogeneous inoculum
on a flat fitness landscape, for the short-range and 
the long-range mean-field 
(all mutations equiprobable) cases.
Since in the real situation  only a small set of the all possible mutations
can occur, we have also considered the quenched version of the long-range 
mutation matrix. In this case
we have shown that a small-world effect is
present, since even a small fraction of quenched long-range
jumps makes the results indistinguishable from those obtained by
assuming all mutations equiprobable. 
These results still hold for a smooth fitness landscape.

We have investigated this issue further, studying the equilibrium
properties of the 
system in the presence of a smooth fitness landscape.
In this framework, it has been possibile to show that
the equilibrium distribution is a Boltzmann one, in which the fitness
plays the role of an energy, and mutations that of a temperature.
We have checked numerically this result
for different fitness landscapes, and a mean-field long-range 
mutation mechanism. 
As in the previous case, a small-world phenomenon appears, 
since similar results
can be obtained using a combination of sparse long-range 
and short-range mutations.

\section*{Acknowledgements}
We  wish to acknowledge our
participation to the DOCS~\cite{DOCS} group.
The numerical simulations have been performed using the 
computational facilities of INFM PAIS 1999-G-IS-Firenze. 

\section*{Appendix: spectral properties of the mean-field mutation
matrices}

Both $\mat{M}_s$ and $\mat{M}_{\ell}$, Eqs.
(\ref{Ms}-\ref{Ml}),  are Markov matrices. 
Moreover, they are circular
matrices, since the value of a given element does not depend on its
absolute position but only on the distance from the diagonal. This
means that their spectrum is real, and that the largest
eigenvalue is $\lambda_{\vec{0}}=1$. Since the matrices are irreducible, the
corresponding  eigenvector $\vec{\xi}_{\vec{0}}$ is non-degenerate, and corresponds
to the flat distribution $\xi_{\vec{0}}(\vec{x}) = 1/2^L$.
In analogy with circular matrices in the usual space, one 
can obtain their complete
spectrum  using the
analogous of Fourier transform in a Boolean hyper-cubic space. Let us
define the ``Boolean scalar product'' $\scalar{}{}$:
\[
	\scalar{\vec{x}}{\vec{y}} = \bigoplus_{i=1}^L x_i y_i,
\]
where the symbol $\oplus$ represent the sum modulus two (XOR) and the
multiplication can be substituted by an AND (which has the same effect
on Boolean quantities). This scalar product is obviously distributive
with respect to the XOR:
\[
	\scalar{(\vec{x} \oplus \vec{y})}{\vec{z}} = (\scalar{\vec{x}}{\vec{z}}) 
	\oplus (\scalar{\vec{y}}{\vec{z}}).
\]
Note that the operation $\vec{x} \oplus \vec{y}$ 
is performed bitwise between the
two genotypes: $(\vec{x} \oplus \vec{y})_i = x_i \oplus y_i$.

Given a function $f(\vec{x})$ of a Boolean quantity 
$\vec{x} \in \{0,1\}^L$, its ``Boolean 
Fourier transform'' is $\tilde f (\vec{k})$ ($\vec{k} \in \{0,1\}^L$)
\[	
	\tilde f(\vec{k}) = \dfrac{1}{2^L} \sum_{\vec{x}} 
		(-1) ^{\scalar{\vec{x}}{\vec{k}}} f(\vec{x}).
\]
The anti-transformation operation is determined by the definition of the
Kronecker delta
\[
	\delta_{\vec{k}\vec{0}} = \dfrac{1}{2^L} \sum_{\vec{x}} 
		(-1) ^{\scalar{\vec{x}}{\vec{k}}},
\]
and is given by
\[
	f(\vec{x}) = \sum_{\vec{k}} (-1) ^ {\scalar{\vec{x}}{\vec{k}}} 
	\tilde f(\vec{k}).
\]

One can easily verify that the Fourier vectors $\xi_k(\vec{x}) =
(-1)^{\scalar{\vec{x}}{\vec{k}}}$ 
are eigenvectors of both $\mat{M}_{\ell}$ and $\mat{M}_s$, 
with eigenvalues
\begin{equation}	
	\begin{array}{rl}
		\lambda_{\vec{0}}&=1,\\
		\lambda_{\vec{k}} &= 1-\mu_{\ell}-\dfrac{\mu_{\ell}}{2^L-1},
	\end{array}
	\label{Mlspectrum}
\end{equation}
for the long-range case, and
\begin{equation}
	\begin{array}{rl}
		\lambda_{\vec{0}}&=1,\\
		\lambda_{\vec{k}} &= 1-\dfrac{2\mu_s d(\vec{k},\vec{0})}{L},
	\end{array}
	\label{Msspectrum}
\end{equation}
for the short-range case, where $d(\vec{x},\vec{y})$ is the Hamming distance
between genotypes $\vec{x}$ and $\vec{y}$. 

The computation of $\rho(t)$, Eq.~(\ref{rho}), 
 is easily performed in Fourier space,
using the analogous of Parsifal theorem:
\[
	\begin{split}
	\sum_{\vec{x}} f(\vec{x}) g(\vec{x}) &= 
	\sum_{\vec{x}} \sum_{\vec{k}} \sum_{\vec{k}'} \tilde f(\vec{k}) \tilde g(\vec{k}')
	(-1)^{\scalar{\vec{x}}{(\vec{k}\oplus \vec{k}')}} \\
	 &= 2^L \sum_{\vec{k}} \sum_{\vec{k}'} \tilde f(\vec{k}) 
	 \tilde g(\vec{k}') \delta_{\vec{kk}'} \\
	 &=	2^L \sum_{\vec{k}}  \tilde f(\vec{k}) \tilde g(\vec{k}),
	 \end{split}
\]
where we have used the property $\vec{k}\oplus \vec{k}'=\vec{0}$ if and only if 
$k_i=k'_i$
for each component $i$.

Let us denote by $\vec{e}_{n}$ the unit vector along direction $n$, i.e.\
$(\vec{e}_{n})_i = \delta_{ni}$.

The Fourier transform of the distance $d(\vec{x},\vec{y})$ 
is obtained considering
that $(-1)^{\scalar{(\vec{x}\oplus \vec{y})}{\vec{e}_n}}$ gives $1$ if $x_n =
y_n$ and $-1$ otherwise, thus
\[
	d(\vec{x},\vec{y}) = \dfrac{1}{2} \left(L-\sum_{n=0}^{L-1}
	(-1)^{\scalar{(\vec{x}\oplus \vec{y})}{\vec{e}_n}}\right).
\]
We obtain 
\[
	\begin{split}
	\tilde d_{\vec{x}_0}(\vec{k}) &= \frac{1}{2^L} \sum_{\vec{x}} 
		d(\vec{x},\vec{x}_0)
			(-1)^{\scalar{\vec{x}}{\vec{k}}}\\
		&= \frac{1}{2^L} \frac{1}{2} \sum_{\vec{x}} \left( L-\sum_{n=0}^{L-1}
	(-1)^{\scalar{(\vec{x}\oplus \vec{x}_0)}{\vec{e}_n}}\right)
			(-1)^{\scalar{\vec{x}}{\vec{k}}}\\
		&= \frac{L}{2} \delta_{\vec{k}\vec{0}} - \frac{1}{2} 
		\left(\sum_{n=0}^{L-1}(-1)^{\scalar{\vec{x}_0}{\vec{e}_n}}\right) \left(
		 \sum_{\vec{x}} (-1)^{\scalar{\vec{x}}{(\vec{k}\oplus \vec{e}_n})}\right)\\
		 &=\frac{L}{2} \delta_{\vec{k}\vec{0}} -\frac{1}{2} \sum_{n=0}^{L-1}
		 (-1)^{\scalar{\vec{x}_0}{\vec{e}_n}} \delta_{\vec{k}\vec{e}_n};
	\end{split}
\]
i.e.
\[	
	\tilde d_{\vec{x}_0}(\vec{k}) = \begin{cases}
			L/2 & \text{if $\vec{k}=\vec{0}$},\\
			-1/2 & \text{if $\vec{k}=\vec{e}_n$ and $(\vec{x}_0)_n=0$},\\
			1/2 & \text{if $\vec{k}=\vec{e}_n$ and $(\vec{x}_0)_n=1$},\\
			0 & \text{otherwise.}
		\end{cases}
\]

The probability distribution $p(\vec{x},t)$ can be expanded on the
eigenvector basis $\xi_k(\vec{x})$ of $\mat{M}$:
\[
	p(\vec{x},0) = \sum_{\vec{k}} a_{\vec{k}} 
		\xi_{\vec{k}}(\vec{x}) = \sum_{\vec{k}} a_{\vec{k}} 
			(-1) ^{\scalar{\vec{x}}{\vec{k}}},
\]
and 
\[
	p(\vec{x},t) = \mat{M}^t p(\vec{x},0) = 
	 \sum_{\vec{k}} a_{\vec{k}} \lambda_{\vec{k}}^t \xi_{\vec{k}}(\vec{x}),
\]
i.e. 
\[
	\tilde p(\vec{k},t) = a_{\vec{k}} \lambda_{\vec{k}}^t.
\]
Thus
\[
	\begin{split}
	\rho(t) &=  \sum_{\vec{k}} \tilde d_{\vec{x}_0}(\vec{k}) \tilde p(\vec{k},t) \\
	 &= \sum_{\vec{k}} \tilde d_{\vec{x}_0}(\vec{k}) a_{\vec{k}} \lambda_{\vec{k}}^t\\
	 &=\dfrac{L}{2} \lambda_{\vec{0}}^t + \sum_{n=0}^{L-1} 
	 \dfrac{1}{2} a_{\vec{e}_n}
	 \lambda_{\vec{e}_n}^t.
	\end{split}
\]

If at $t=0$ the distribution is concentrated at $\vec{x}_0=\vec{0}$
($p(\vec{x},0)=\delta_{\vec{x}\vec{0}}$) then $a_{\vec{k}}=1$ for all $\vec{k}$. 
In both the short and long-range cases, $\lambda_{\vec{e}_n}$ does not
depend on $n$ (see Eqs.~\eqref{Mlspectrum}-\eqref{Msspectrum}), and thus 
\[
	\rho(t) = \dfrac{L}{2} \left(\lambda_{\vec{0}}^t -\lambda_{\vec{1}}^t\right).
\]

For the short-range case we have 
\[
	\begin{split}
	\rho_s(t) &= \dfrac{L}{2} 
		\left(1 -\left(1-\dfrac{2\mu_s}{L}\right)^t\right)\\
		&= \dfrac{L}{2} \left(1 -\exp\left(t
		\ln\left(1-\dfrac{2\mu_s}{L}\right)\right)\right) \\
		&\simeq
		\dfrac{L}{2}\left(1-\exp\left(-\dfrac{t}{\tau_s}\right)\right),
	\end{split}
\]
and the characteristic spreading time is 
$\tau_s=L/2\mu_s$. For $t$ small compared with $\tau_s$
($L\rightarrow \infty$) we have
\begin{equation}\label{vs}
	\rho_s(t) \simeq \mu_s t \equiv v_s t.
\end{equation}

For long-range mutations we have 
\[
	\begin{split}
	\rho_{\ell}(t) &= \dfrac{L}{2} \left(1 -\left(1-\mu_\ell -
	\dfrac{\mu_{\ell}}{\vec{e}_nL-1}\right)^t\right)\\
		&\simeq	
\dfrac{L}{2}\left(1-\exp\left(-\dfrac{t}{\tau_\ell}\right)\right)
	\end{split}
\]
with $\tau_\ell=1/\mu_\ell$. For $t<<\tau_\ell$
($\mu_\ell\rightarrow 0$) we have
\begin{equation}\label{vl}
	\rho_{\ell}(t) \simeq \dfrac{L\mu_\ell}{2} t \equiv v_{\ell} t.
\end{equation}

The behavior of $\rho(t)$ for short times, vanishing mutation
probabilities and large genotypes, Eqs.~\eqref{vs}-\eqref{vl}, is
rather trivial. In these approximations one can neglect back mutations, 
and obtain the same results on a Cayley tree. However, 
the full analysis gives the exact behavior of $\rho(t)$ for all times.

In the mixed case one has
$\mat{M}=\mat{M}_s\mat{M}_\ell$. Since 
$\mat{M}_s$ and $\mat{M}_\ell$ share the same
eigenvectors, $\lambda=\lambda_s \lambda_\ell$ and, in the previous
approximations,
\begin{equation}\label{v}
		v = v_s + v_\ell. 
\end{equation}


\newpage


\begin{table}
\begin{tabular}{|c||c|c|c||c|c|c||c|c|c||c|c|c|}
$V$   &  
\multicolumn{3}{c||}{$L=6$} & \multicolumn{3}{c||}{$L=8$}
& \multicolumn{3}{c|}{$L=10$}
& \multicolumn{3}{c|}{$L=12$} \\ 
	\hline
  & $\mu_e^{(0)}$ & $\overline{\mu}_e$ & $s_c$ &
$\mu_e^{(0)}$ & $\overline{\mu}_e$ & $s_c$ &	
$\mu_e^{(0)}$ & $\overline{\mu}_e$ & $s_c$ &
$\mu_e^{(0)}$ & $\overline{\mu}_e$ & $s_c$ \\
\hline
$\mcal{H}$& 0.035 & 0.045 & 3 & 0.025 & 0.035 & 3 & 0.020 & 0.030 & 2 
	& 0.017 & 0.027& 2\\
\hline				
$\mcal{J}$ & 0.071 & 0.082 & 4 & 0.053 & 0.063 & 4 & 0.042 & 0.052 & 4 
	& 0.035 & 0.045 & 3\\
\hline				
$\mcal{K}$  & 0.080 & 0.105 & 6& 0.090 & 0.105 & 6 & 0.096 & 0.107 & 6 
	& 0.098 & 0.111& 3\\
\end{tabular}
\caption{\label{table:s} Effettive temperature $\mu_e$ for several values 
of the  genotype length $L$ and parameters of the fitness $V(\vec{x})$,
Eq~(\protect\ref{fitness}). Here  $\mu_s=0.1$, $\mu_{\ell}=0.01$; the row 
labeled  $\mcal{H}$ stands for $\mcal{H}=0.1$, $\mcal{J}=\mcal{K}=0$,
the row 
labeled  $\mcal{J}$ stands for $\mcal{J}=0.1$, $\mcal{H}=\mcal{K}=0$,
the row 
labeled  $\mcal{K}$ stands for $\mcal{K}=0.1$, $\mcal{H}=\mcal{J}=0$.
We report the value or $\mu_e^{(0)}$ for 
sparseness $s=0$ (only small-range mutations) and the average value 
$\overline{\mu}_e$
for $s>s_c$,
where $s_c$ is estimated visually from the plot of data 
(see also Fig.~\protect\ref{figure:sparsechi}). 
}
\end{table}

\end{document}